\documentclass[aps,floatfix,superscriptaddress,reprint,10pt,pra]{revtex4-1}
\usepackage{bbm}
\usepackage{bm}
\usepackage{amsmath}
\usepackage{amssymb}
\usepackage{empheq}
\usepackage{graphicx}
\usepackage{mathrsfs}
\usepackage{amsfonts}
\usepackage{amsthm}
\usepackage{color}
\usepackage{bigints}
\usepackage{txfonts}
\usepackage{hyperref}
\usepackage{color}
\usepackage{appendix}
\usepackage{multirow}
\usepackage{makecell}

\hypersetup{
	unicode=false,               pdftoolbar=true,             pdfmenubar=true,             pdffitwindow=false,          pdfstartview={FitH},         pdftitle={My title},         pdfauthor={Author},          pdfsubject={Subject},        pdfcreator={Creator},        pdfproducer={Producer},      pdfkeywords={keyword1} {key2} {key3},      pdfnewwindow=true,           colorlinks=false,            linkcolor=red,               citecolor=green,             filecolor=magenta,           urlcolor=cyan           }
\makeatletter \tolerance = 10000 \tolerance = 10000
\makeatother

\begin{document}
	\title{Strain Engineering of Intrinsic Anomalous Hall and Nernst Effects in Altermagnetic MnTe at Realistic Doping Levels}
	
	\author{Weiwei Chen}
	\affiliation{Department of Physics, Hefei University of Technology, Hefei, Anhui 230601, China}
	
	\author{Ziyu Zhou}
	\affiliation{Department of Physics, Hefei University of Technology, Hefei, Anhui 230601, China}
	
	\author{Jie Meng}
	\affiliation{College of Chemistry and Materials, Taiyuan Normal University, Jinzhong 030619, China}
	
	\author{Weiyi Wang}
	\affiliation{Hefei National Research Center for Physical Sciences at the Microscale , University of Science and Technology of China, Hefei, Anhui 230026, China.}
	
	\author{Ye Yang}
	\thanks{Contact author. E-mail: yangye@hfut.edu.cn}
	\affiliation{Department of Physics, Hefei University of Technology, Hefei, Anhui 230601, China}
	
	\author{Zhongjun Li}
	\thanks{Contact author. E-mail: zjli@hfut.edu.cn}
	\affiliation{Department of Physics, Hefei University of Technology, Hefei, Anhui 230601, China}
\begin{abstract}
Hexagonal MnTe has emerged as a prototypical g-wave altermagnet, hosting time-reversal symmetry breaking in momentum space despite a vanishing net magnetization. While this symmetry breaking theoretically allows for an intrinsic anomalous Hall effect, experimentally observed signals have remained weak. In this work, we investigate the origin of this suppression and demonstrate a strategy to amplify anomalous transport responses within the experimentally accessible doping regime. Using a $\bm{k}\cdot\bm{p}$ effective model, we reveal that near the valence band maximum, which corresponds to the energy window relevant for typical hole doping ($\sim10^{19}cm^{-3}$), the intrinsic Hall effect is suppressed due to a symmetry-enforced cancellation of opposing Berry curvature contributions. We propose that breaking the crystalline symmetry via volume-conserving biaxial strain lifts this cancellation, resulting in a significant enhancement of the anomalous Hall conductivity by orders of magnitude. This strain-induced Fermi surface distortion also  amplifies the anomalous Nernst effect. Furthermore, the analysis of the spin texture confirms that these strain-enabled anomalous transport signatures emerge while preserving the zero net magnetization. 
\end{abstract}

\date{\today}

\maketitle

\section{Introduction}

Following the proposal of altermagnetism\cite{Smejkal2022prx2}, first-principle calculations identified hexagonal MnTe as a primary candidate hosting g-wave altermagnetic order \cite{Betancourt2023prl,Lee2024prl,Rooj2023apr}. This prediction has been corroborated by a broad range of experiments, including transport measurements \cite{Betancourt2023prl,Lee2024prl,Kluczuyk2024prb,Chicote2024afm}, spectroscopic probes \cite{Hariki2024prl,Osumi2024prb,Lee2024prl}, and direct real-space nanoscale imaging \cite{Amin2024nature}. A fundamental characteristic of altermagnets is that they break time-reversal symmetry in momentum space while maintaining zero net magnetization \cite{Smejkal2022prx,Fedchenko2024sa}. When combined with spin-orbit coupling, this symmetry breaking enriches the spin texture in momentum space \cite{Chen2025cpl} and allows for the emergence of the anomalous Hall effect \cite{Sato2024prl,Attias2024prb,Roig2025prl}, a phenomenon typically associated with ferromagnets.

Despite these theoretical expectations, initial experimental observations in $p$-type doped MnTe have reported weak AHE signals, with anomalous Hall conductivity reaching only $\sim0.02S/cm$ \cite{Betancourt2023prl,Takahashi2025npj}. This discrepancy has sparked intense debate regarding the origin of the signal. Specifically, whether the observed weak response arises from intrinsic altermagnetism or from parasitic ferromagnetism due to canted moments \cite{Smolenski2025arxiv} or stoichiometry-induced defects \cite{Chicote2024afm}. Theoretical investigations suggest that the negligible anomalous Hall effect signal detected in experiments is intrinsic to the bulk band structure of hexagonal MnTe \cite{Kluczuyk2024prb}. Since typical samples in both bulk and epitaxial layers exhibit $p$-type conductivity \cite{Junior2023prb,Ren2017jmcc,Xin2019ne}, the transport properties are dominated by states near the valence band maximum (VBM). Although crystal and magnetic symmetries permit a non-zero anomalous Hall conductivity, calculations show that the magnitude of the anomalous Hall conductivity is extremely small in this region. The energy states contributing to a strong anomalous response are located significantly deep in the valence band \cite{Devaraj2024prm,Betancourt2023prl}, far from the Fermi level, rendering the large anomalous Hall effect inaccessible at typical doping concentrations ($\sim10^{19}cm^{-3}$). Experimentally, the situation is further complicated by the multi-domain nature of real crystals. Recent studies on MnTe indicate that the observed anomalous Hall effect often coexists with a weak net magnetization arising from domain imbalance \cite{Kluczuyk2024prb,Liu2025arvix}. Consequently, distinguishing the intrinsic altermagnetic anomalous Hall effect from contributions due to weak ferromagnetism or domain distribution artifacts remains a challenge. Addressing this complexity, recent magnetotransport studies on epitaxial films have demonstrated that the anomalous Hall effect signal is strongly anisotropic, governed by the critical interplay between the altermagnetic N\'eel vector and crystal symmetry \cite{Bangar2025arxiv}. This suggests that precise control of crystallographic orientation is essential for isolating the intrinsic altermagnetic transport signature.

Motivated by these developments, we investigate the origin of the suppressed AHE near the VBM using a $\bm{k}\cdot\bm{p}$ model. We focus specifically on the Berry curvature distribution near the $A$ point of the Brillouin zone. Our results reveal that the system actually exists strong local Berry curvature polarization with opposite signs along the $k_x$ and $k_y$ directions. However, upon integration over the hole-occupied states, the total anomalous Hall conductivity becomes negligible. We attribute this cancellation to the underlying $C_6$ rotational symmetry of the Fermi surface, which enforces a compensation between the opposing Berry curvature contributions.

To overcome this symmetry-imposed limitation and enhance the signal near the Fermi level, we propose a strategy to lower the symmetry of the Fermi surface by introducing a volume-conserving biaxial strain to the crystal lattice \cite{Belashchenko2025prl}. Our calculations demonstrate that such strain effectively distorts the Fermi surface, thereby preventing the cancellation of the Berry curvature. This results in a significant enhancement of the anomalous Hall conductivity in MnTe, which exhibits a linear increase with strain strength at low doping concentrations. Furthermore, we show that the anomalous Nernst conductivity is also remarkably amplified by this Fermi surface distortion. Crucially, by analyzing the spin polarization in momentum space under such strain effect, we confirm that the zero net magnetization is preserved. Our findings suggests that the zero-magnetization anomalous Hall and Nernst effect signals in MnTe can be experimentally detected and manipulated by introducing appropriate strain through substrate engineering.

\section{Model and Method}

\begin{figure}
	\centering
	\includegraphics[width=1\linewidth]{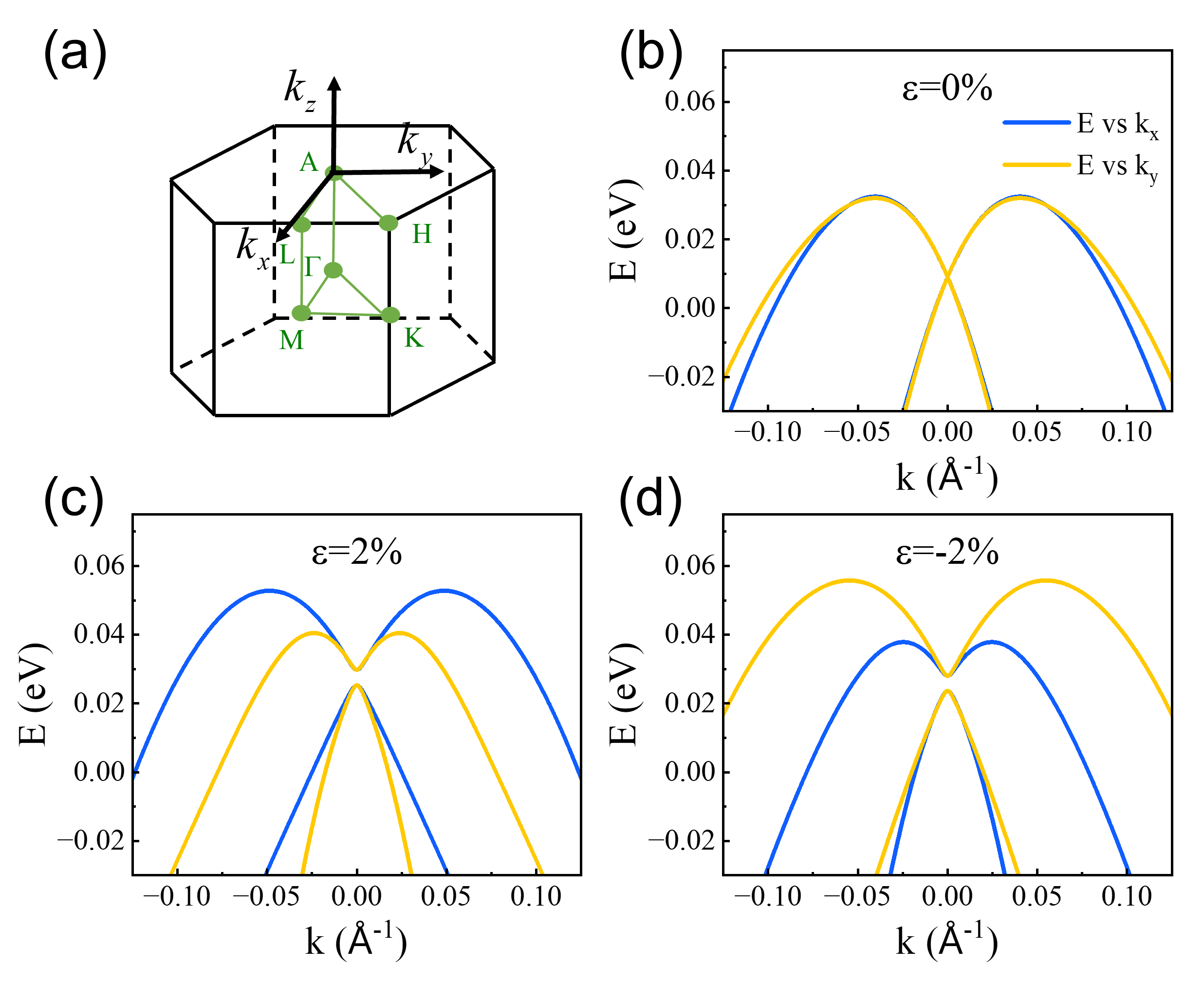}
	\caption{(Color online) (a) Brillouin zone of the hexagonal MnTe. (b-d) Band structures near the valence band maximum (A point) under different strain strengths: (b) $\varepsilon=0\%$, (c) $\varepsilon=2\%$, and (d) $\varepsilon=-2\%$. The blue curves denote the energy dispersion along the $k_x$ (with $k_y=0$), while the yellow curves denote the dispersion along the $k_y$ (with $k_x=0$). All curves are taken at a fixed $k_z$=0.1$\AA^{-1}$.}
	\label{fig:band}
\end{figure}

\begin{figure}
	\centering
	\includegraphics[width=1.0\linewidth]{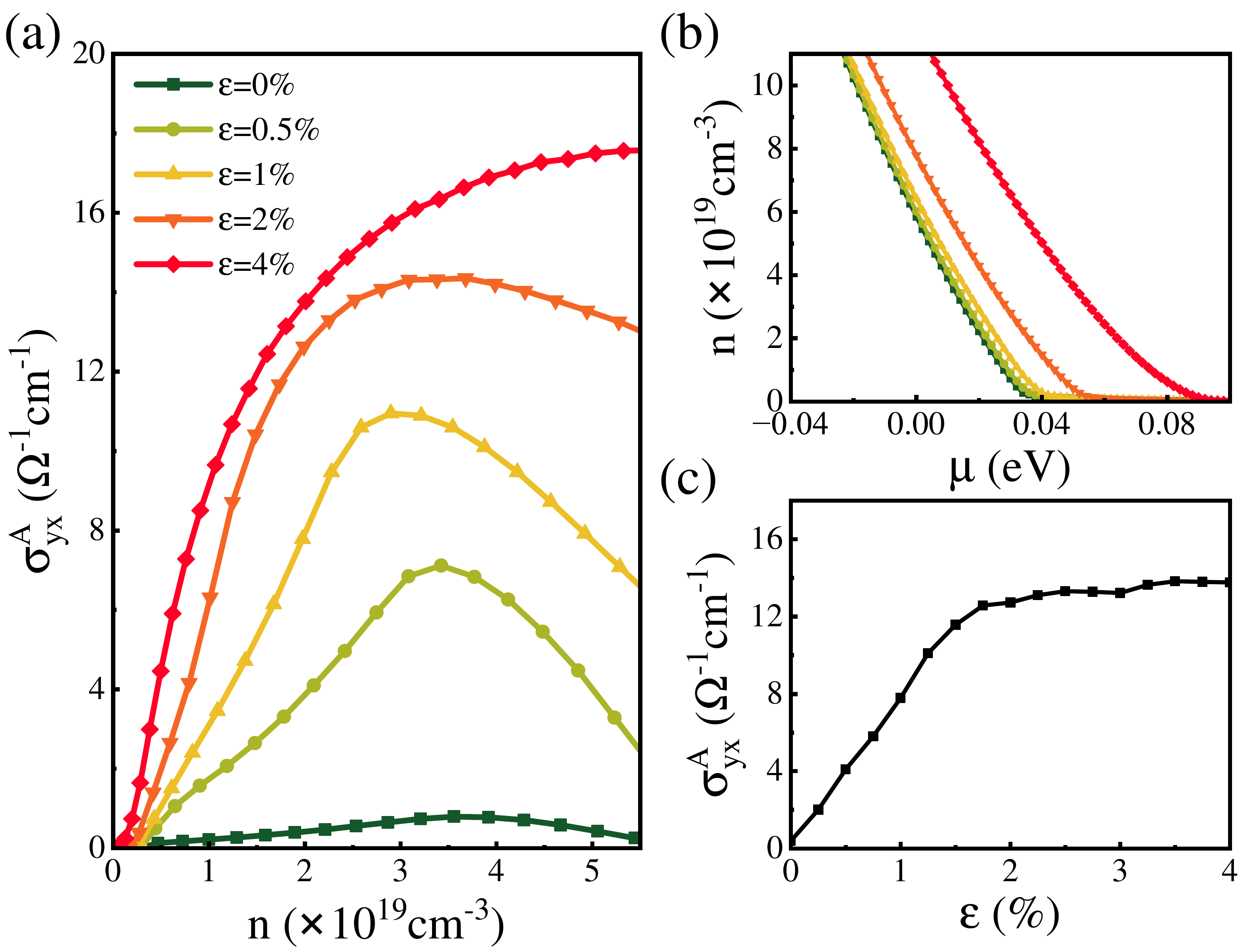}
	\caption{(Color online) (a) Intrinsic anomalous Hall conductivity $\sigma^A_{yx}$ as a function of hole doping concentration $n$ for various strain strengths. (b) The correspondence between hole doping concentration $n$ and chemical potential $\mu$. (c) Strain dependence of $\sigma^A_{yx}$ for a fixed hole concentration of $n=2\times10^{19} cm^{-3}$. Broaden parameter: $\eta=0.01$ eV.}
	\label{fig:ahe-strain}
\end{figure}

We employ the $\bm{k}\cdot \bm{p}$  effective Hamiltonian near the VBM of hexagonal MnTe, as derived in Refs.~\cite{Takahashi2025npj,Belashchenko2025prl,Junior2023prb}. Considering a N\'eel vector aligned with the magnetic easy axis (the $[1\bar{1}00]$ direction) and including spin-orbit coupling, the Hamiltonian in the vicinity of the $A$ point is given by
\begin{equation}\label{eq:Hamiltonian}
	H=H_0+H_{J}+H_{\lambda}+H_{\Delta}
\end{equation}
with
\begin{equation}\begin{aligned}
	H_0=&c_1(k_x^2+k_y^2)+c_2k_z^2+c_3\left[(k_x^2-k_y^2)\tau_z+2k_xk_y\tau_x\right]\\
	H_J=&t_{J}k_z\omega_z(k_x\tau_x+k_y\tau_z)\\
	H_{\lambda}=&\lambda_1(k_x\tau_z-k_y\tau_x)\omega_y+\lambda_2 k_z\tau_z\omega_x
	+\lambda_3k_z\tau_y\omega_x+\lambda_4k_x\tau_y\omega_y\\
	H_{\Delta}=&(\Delta_0+\Delta_s)\tau_z
\end{aligned}\end{equation}
Here, we retain only the leading-order contributions to $H_J$ and $H_{\lambda}$. The basis is defined by the orbital doublet formed by the Te $p_x$ and $p_y$ orbitals. The Pauli matrices $\tau_i$ act on this orbital subspace, while $\omega_i$ represent the pseudospin degrees of freedom arising from the coupling spin and sublattice through the relations: $\omega_x=\sigma_x\nu_x$, $\omega_y=\sigma_y\nu_x$, and $\omega_z=\sigma_z$, where $\sigma_i$ and $\nu$ are Pauli matrices for spin and sublattice.  

The term $H_J$ encodes the altermagnetic spin-dependent hybridization, while $H_{\lambda}$ describes the spin-orbit coupling. The term $H_{\Delta}$ accounts for the orthorhombic distortion of the charge density. This distortion includes an intrinsic contribution, characterized by a quite small value $\Delta_0=1.3$ meV, and a tunable component $\Delta_s$ induced by strain. This strain-induced orbital splitting $\Delta_s$ is derived from the lattice distortion. Following the fitting of first-principle calculations reported in Ref.~\cite{Belashchenko2025prl}, we adopt a linear coupling model:
\begin{equation}
	\Delta_s=2.13\varepsilon \ (\text{eV})
\end{equation}
where $\varepsilon$ represents a volume-conserving biaxial strain defined by the tensor components $\varepsilon_{xx}=-\varepsilon_{yy}=\varepsilon/2$. Although both the $\Gamma$ and $A$ points are located near the VBM \cite{Junior2023prb,Takahashi2025npj}, the effective Hamiltonian near the $\Gamma$ point (dominated by the single $p_z$ orbital of Te) is pure real-valued. Consequently, it yields a vanishing Berry curvature and zero anomalous Hall conductivity. Therefore, our analysis focuses exclusively on the $A$ point. 

\begin{figure*}[t]
	\centering
	\includegraphics[width=1\linewidth]{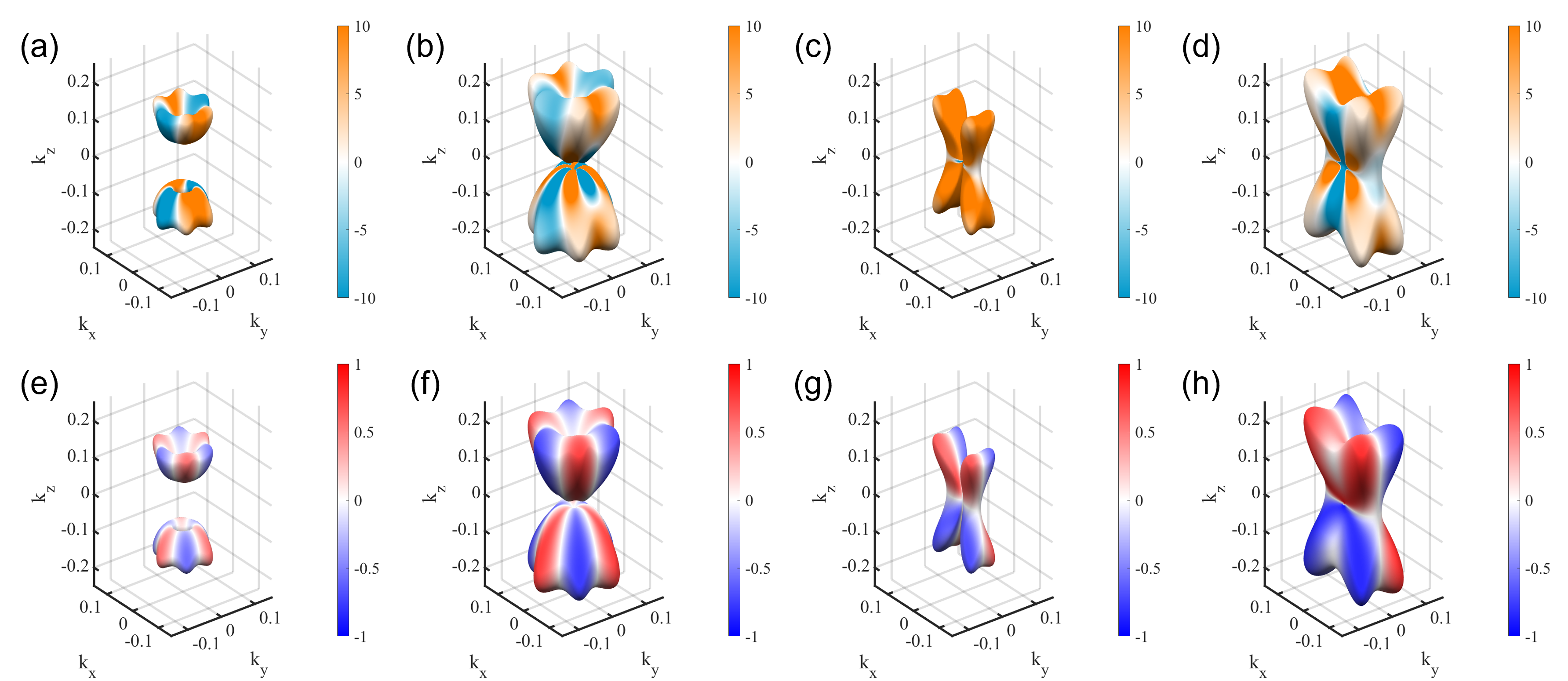}
	\caption{(Color online) Evolution of the Fermi surface, Berry curvature, and spin polarization under strain: $\varepsilon=0\%$ (a)(b)(e)(f) and $\varepsilon=2\%$ (c)(d)(e)(f). The isosurfaces are plotted for hole doping concentrations of $n=1\times10^{19}cm^{-3}$ [(a)(c)(e)(g)] and $4\times10^{19}cm^{-3}$ [(b)(d)(f)(h)]. The colormap in (a-d) depicts the Berry curvature $\Omega_{yx}$ calculated via Eq~(\ref{eq:Berry-curvature}), while (e-h) shows the spin expectation value obtained by $\langle s\rangle=\langle u_{\bm{k}n}|\omega_z|u_{\bm{k}n}\rangle$.}
	\label{fig:iso-strain}
\end{figure*}

The evolution of the band structure near the valence band maximum (A point) under applied strain is presented in Fig.~\ref{fig:band}. In the absence of strain, the energy dispersion along $k_x$ and $k_y$ directions are distinct but degenerate at the extremum. Upon the application of positive strain, this degeneracy is lifted, creating a clear level distinction: the dispersion along $k_x$ shifts to higher energies relative to $k_y$; conversely, this hierarchy is reversed under negative strain. This anisotropy confirms that strain induces a significant orthorhombic distortion near the band edge, and inspires our investigation into the evolution of the Berry curvature distribution.

Using linear response theory, the charge current density $\bm{J}$ induced by electric field $\bm{E}$ and temperature gradient $\nabla T$ is given by
\begin{equation}
	J_j=\sigma_{ji}E_i+\alpha_{ji}(-\nabla_i T)
\end{equation}
where $\sigma_{ji}$ and $\alpha_{ji}$ denote the electrical and thermoelectric conductivity tensors, respectively.  The intrinsic anomalous Hall conductivity $\sigma^A_{yx}$, which characterizes the response of transverse current to the longitudinal electric field, can be evaluated by \cite{Zeng2006prl,Nagaosa2010rmp,Takahashi2017prb}
\begin{equation}\label{eq:AHC}
	\sigma^A_{yx}=\frac{e^2}{\hbar}\sum_{n}\int\frac{d^d\bm{k}}{(2\pi)^d}\Omega_{yx}^{n}(\bm{k})f_h(\tilde{E}_{n\bm{k}},\tilde{\mu})
\end{equation}
where $\Omega_{yx}^n$ is the Berry curvature defined as
\begin{equation}\begin{aligned}\label{eq:Berry-curvature}
		\Omega_{yx}^n(\bm{k})=i\sum_m\left[\frac{\langle u_{\bm{k}n}|v_y|u_{\bm{k}m}\rangle\langle u_{\bm{k}m}|v_x|u_{\bm{k}n}\rangle-(v_y\leftrightarrow v_x)}{(E_{\bm{k}n}-E_{\bm{k}m})^2}\right]
\end{aligned}\end{equation}
with $v_i=\frac{\partial H}{\hbar\partial k_i}$ the velocity operators and $u_{\bm{k}n}$ eigen wavefunction. In the numerical calculations, to regularize the divergence arising from degenerate states, we introduce a finite broadening parameter $\eta$ in the denominator \cite{Yates2007prb}, i.e. $(E_{\bm{k}n}-E_{\bm{k}m})^2\to(E_{\bm{k}n}-E_{\bm{k}m})^2+\eta^2$. $f_h$ is the Fermi-Dirac distribution function for holes, defined as
\begin{equation}
	f_h(\tilde{E}_{n\bm{k}}, \tilde{\mu}) = \frac{1}{e^{(\tilde{E}_{n\bm{k}}-\tilde{\mu})/k_B T}+1}.
\end{equation}
with the hole energy $\tilde{E}_{n\bm{k}}$ and hole chemical potential $\mu$ defined by
\begin{equation}
	\tilde{E}_{n\bm{k}}=E_{V}-E_{n\bm{k}},\ \ \tilde{\mu}=E_{V}-\mu
\end{equation}
where $E_V$ denotes the valence band maximum.

Similarly, the anomalous Nernst conductivity, which characterizes the response of transverse current to the longitudinal temperature gradient, can be evaluated by \cite{Xiao2006prl,Bergman2010prl}
\begin{equation}\label{eq:ANC}
	\alpha^{A}_{yx}=-\frac{ek_B}{\hbar}\sum_{n}\int\frac{d^d\bm{k}}{(2\pi)^d}\Omega_{yx}^{n}(\bm{k})s_h(\tilde{E}_{n\bm{k}},\tilde{\mu})
\end{equation}
where $s_h$ denotes the entropy density of the hole states, given by
\begin{equation}
	s_h=-f_h\ln f_h-(1-f_h)\ln(1-f_h).
\end{equation}

\section{Anomalous Hall Effects}

Figure~\ref{fig:ahe-strain}(a) presents the calculated intrinsic AHC, $\sigma^A_{yx}$, obtained from Eq.~(\ref{eq:AHC}) with zero temperature, as a function of hole doping concentration $n$ for various strain strengths. The correspondence between the doping concentration $n$ and the chemical potential $\mu$ is provided in Fig.~\ref{fig:ahe-strain}(b). By comparing these correspondences with the band dispersion near the VBM shown in Fig.~\ref{fig:band}, it is evident that the typical doping concentration ($\sim10^{19}cm^{-3}$) corresponds to a narrow energy window dominated by the topmost valence band. Our calculations indicate that strain serves as a powerful control knob for the AHE in this system within typical doping ranges. In the absence of strain ($\varepsilon=0\%$), $\sigma^A_{yx}$ is relatively small near the valence band edge. However, the application of tensile strain significantly enhances the AHC. As detailed in Fig.~\ref{fig:ahe-strain}(c) for a fixed concentration $n=2\times10^{19}cm^{-3}$, the $\sigma^A_{yx}$ exhibits a linear increase with strain strength up to $\varepsilon\approx2\%$, beyond which it exhibits saturation behavior.

To elucidate the microscopic origin of this giant strain response, we analyze the distribution of the Berry curvature $\Omega_{yx}(\bm{k})$ on the Fermi surfaces, as shown in Figs.~\ref{fig:iso-strain}(a-d). In this analysis, we neglect the intrinsic splitting $\Delta_0$, as it produces effects qualitatively similar to strain but is negligible in magnitude compared to the strain-induced $\Delta_s$. In the pristine limit ($\varepsilon=0\%$), the Fermi surface exhibits a $C_6$ rotational symmetry consistent with the hexagonal lattice [Figs.~\ref{fig:iso-strain}(a) and (b)]. The Berry curvature distribution, however, reflects a lower effective symmetry due to the interplay of altermagnetism and spin-orbit coupling. Regions of positive and negative Berry curvature are concentrated along the $k_x$ and $k_y$ directions, respectively. The integration of $\Omega_{yx}$ over the Fermi surfaces is subject to substantial cancellation between these opposing contributions, resulting in a suppressed net anomalous Hall conductivity. 

The application of strain induces the $p_x/p_y$ orbital hybridization via the $\Delta_s$ term. This distortion deforms the Fermi surface and, more crucially, disrupts the delicate cancellation of the Berry curvature. As shown in Fig.~\ref{fig:iso-strain}(c) for the low-doping case ($n=1\times10^{19}cm^{-3}$) under $2\%$ strain, the regions of negative Berry curvature are almost completely suppressed while positive contributions dominate, driving the large enhancement of the macroscopic AHC observed in Fig.~\ref{fig:ahe-strain}. At higher doping concentration, the Fermi surface expands and multi-band effects become important. This expansion tends to mitigate the strain-induced distortion, regenerating a partial cancellation between positive and negative Berry curvature regions, as shown in Fig.~\ref{fig:iso-strain}(d). As a result, the intrinsic AHC does not increase monotonically with $n$, consistent with the trends shown in Fig.~\ref{fig:ahe-strain}(a).

Furthermore, we have compared the spin expectation values $\langle s\rangle=\langle u_{\bm{k}n}|\omega_z|u_{\bm{k}n}\rangle$ projected onto the isosurface in the absence and presence of strain, as shown in Figs.~\ref{fig:iso-strain}(e-h). In the absence of strain, the spin texture exhibits a g-wave polarization pattern, a characteristic signature of the altermagnetism in this system. Remarkably, even under strain, the spin texture retains an antisymmetry that ensures the net macroscopic magnetization remains zero ($\sum_{\bm{k}}\langle s\rangle=0$). This confirms that the strain-induced enhancement of the AHE arises purely from the modification of band topology and Berry curvature, rather than from an induced ferromagnetic moment.

\section{Anomalous Nernst Effects}

\begin{figure}
	\centering
	\includegraphics[width=1\linewidth]{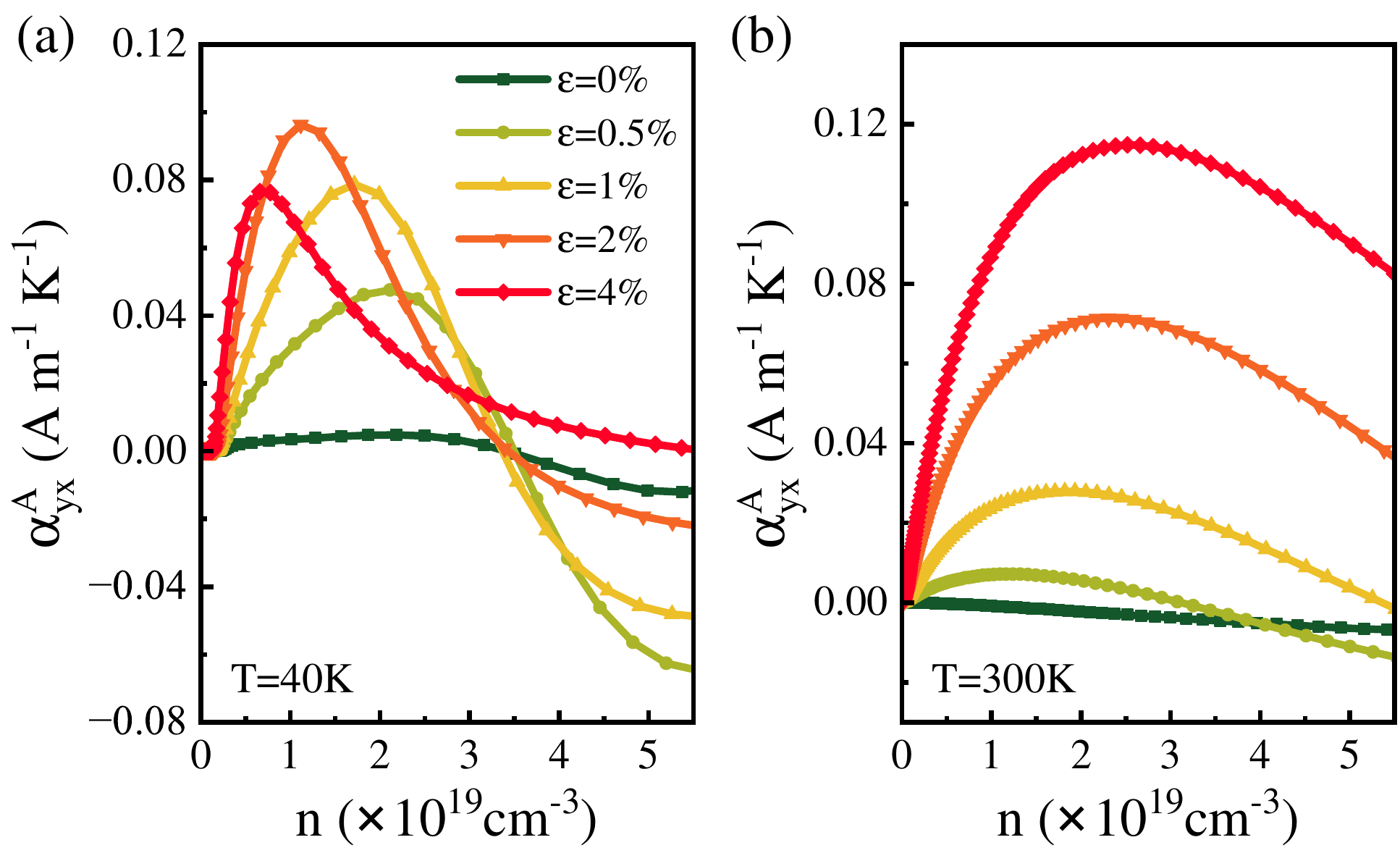}
	\caption{(Color online) Intrinsic anomalous Nernst conductivity $\alpha^A_{yx}$ as a function of hole doping concentration $n$ for various strain strengths $\varepsilon$ at temperatures of (a) $T=40K$ and (b) $T=300K$.}
	\label{fig:ane-strain}
\end{figure}

Figure~\ref{fig:ane-strain} presents the calculated intrinsic anomalous Nernst conductivity, $\alpha^A_{yx}$, determined via Eq.~(\ref{eq:ANC}), as a function of hole doping concentration $n$ for various strain strengths. Similar to the behavior observed in the anomalous Hall conductivity, the symmetry breaking induced by orthorhombic distortion significantly amplifies the anomalous Nernst conductivity near the valence band edge.

At low temperature ($T=40K$), the anomalous Nernst conductivity exhibits a sharp increase followed by a rapid decrease, notably undergoing a sign reversal from positive to negative at a doping concentration of approximately $n\approx3\times10^{19}cm^{-3}$ under strain, as shown in Fig.~\ref{fig:ane-strain}(a). This non-monotonic behavior can be understood via the Mott relation, which links the thermoelectric response to the energy derivative of the electrical conductivity at low temperatures \cite{Xiao2006prl,Behnia2016rpp}:
\begin{equation}
	\alpha_{ji}=\frac{\pi}{3}\frac{k_B^2T}{e}\frac{d\sigma_{ji}}{d\epsilon}\bigg|_{\epsilon=\mu}.
\end{equation}
According to this relation, the magnitude of $\alpha^A_{yx}$ is proportional to the slope of $\sigma^A_{yx}$ with respect to the chemical potential. Consequently, the peaks in the Nernst signal correspond to the regions where $\alpha_{yx}^A$ changes most rapidly. The observed sign reversal indicates a transition point where the anomalous Hall conductivity reaches a local maximum \cite{Alam2023prb}, reflecting the complex evolution of the Berry curvature integration as the Fermi surface expands to encompass multiple bands.

In contrast to the complex dependence at low temperature, the behavior at room temperature ($T=300K$) is more uniform. As shown in Fig.~\ref{fig:ane-strain}(b), $\alpha_{yx}^A$ increase monotonically with strain $\varepsilon$ across a wide range of doping concentrations ($n\in[0,4]\times10^{19}cm^{-3}$).

\section{Summary and Discussion}
In summary, we have elucidated the microscopic origin of the suppressed anomalous transport in hexagonal MnTe near the VBM and demonstrated a strain-engineering strategy to unlock its topological potential. Based on the $\bm{k}\cdot\bm{p}$ effective model focused on the experimentally relevant doping concentration ($\sim10^{19}cm^{-3}$), we identified that the weak intrinsic anomalous Hall effect arises from a symmetry-enforced cancellation of opposing curvature contributions in momentum space. We show that breaking the crystalline symmetry via volume-conserving biaxial strain effectively lifts this cancellation.

Our results demonstrate that a moderate lattice distortion ($\sim2\%$) deforms the Fermi surface to  sufficiently enhance the anomalous Hall conductivity by orders of magnitude. In the meanwhile, we predict a simultaneous amplification of the anomalous Nernst conductivity, which exhibits a monotonic behavior with doping concentration and exists a sign reversal at approximately $n\approx3\times10^{19}cm^{-3}$.

The strategy we propose is timely, as recent experiments have demonstrated the exceptional tunability of the anomalous Hall effect in MnTe under stress \cite{Smolenski2025arxiv,Liu2025arvix}. Our $\bm{k}\cdot\bm{p}$ results align with the ``piezo-Hall" mechanism proposed by Smolenski et al. \cite{Smolenski2025arxiv}, where strain directly modifies the multipolar Berry curvature distribution. Furthermore, our analysis suggests that the strain-induced enhancement persists even when considering the domain detwining effects reported by Liu et al. \cite{Liu2025arvix}, as the intrinsic conductivity of a single domain is fundamentally amplified by the symmetry breaking. This is further supported by the work of Bangar et al. \cite{Bangar2025arxiv}, who showed that the transport response is highly sensitive to the relative orientation of the crystal axes, reinforcing the necessity of crystal-symmetry control.

Finally, both our calculations and the experimental magnetometry data confirm that the net magnetization remains negligible under strain. This rules out piezomagnetism (induced ferromagnetism) as the primary driver and firmly establishes the response as intrinsic to the altermagnetism band topology. Our findings not only clarify the mechanism behind the strain-tunable anomalous effect but also suggest that anomalous Nernst effect should be a prime target for future experimental verification in strained MnTe devices.

\textit{Acknowledgments.---}	
This work is supported by the National Science Foundation of China No. 12574048 (YY), No. 12174080 (ZL), and No. 12547210 (ZL), the China National Key Research and Development Program Grant No. 2022YFA1602601 (ZL), and the Fundamental Research Funds for the Central Universities No. JZ2025HGQA0310 (WC).


\begin{thebibliography}{99}
	
\bibitem{Smejkal2022prx2} L. \u{S}mejkal, J. Sinova, and T. Jungwirth,
Beyond Conventional Ferromagnetism and Antiferromagnetism: A Phase with Nonrelativistic Spin and Crystal Rotation Symmetry,
\href{https://doi.org/10.1103/PhysRevX.12.031042}
{Phys. Rev. X \textbf{12}, 031042 (2022).}

\bibitem{Rooj2023apr}
S. Rooj, J. Chakraborty, N. Ganguli,
Hexagonal MnTe with Antiferromagnetic Spin Splitting and Hidden Rashba–Dresselhaus Interaction for Antiferromagnetic Spintronics,
\href{https://doi.org/10.1002/apxr.202300050}
{Adv. Physics Res. \textbf{3}, 2300050 (2024).}

\bibitem{Betancourt2023prl} R. D. G. Betancourt, J. Zub\'a\v{c}, R. Gonzalez-Hernandez, K. Geishendorf, Z. \v{S}ob\'a\v{n}, G. Springholz, K. Olejn\'{i}k, L. \v{S}mejkal, J. Sinova, T. Jungwirth, S. T. B. Goennenwein, A. Thomas, H. Reichlov\'a, J. \v{Z}elezn\'y,2 and D. Kriegner,
Spontaneous Anomalous Hall Effect Arising from an Unconventional Compensated Magnetic Phase in a Semiconductor
\href{https://doi.org/10.1103/PhysRevLett.130.036702}
{Phys. Rev. Lett. \textbf{130}, 036702 (2023).}

\bibitem{Lee2024prl} Suyoung Lee, Sangjae Lee, S. Jung, J. Jung, D. Kim, Y. Lee, B. Seok, J. Kim, B. G. Park, L. \v{S}mejkal, C-J. Kang, and C. Kim
Broken Kramers Degeneracy in Altermagnetic MnTe,
\href{https://doi.org/10.1103/PhysRevLett.132.036702}
{Phys. Rev. Lett. \textbf{132}, 036702 (2024).}

\bibitem{Kluczuyk2024prb} K. P. Kluczyk, K. Gas, M. J. Grzybowski, P. Skupi\'nski, M. A. Borysiewicz, T. Fas, J. Suffczy\'nski, J. Z. Domagala, K. Grasza, A. Mycielski, M. Baj, K. H. Ahn, K. V\'yborn\'y, M. Sawicki, and M. G-Borysiewicz,
Coexistence of anomalous Hall effect and weak magnetization in a nominally collinear antiferromagnet MnTe,
\href{https://doi.org/10.1103/PhysRevB.110.155201}
{Phys. Rev. B \textbf{110}, 155201 (2024).}

\bibitem{Chicote2024afm}
M. Chilcote, A. R. Mazza, Q. Lu, I. Gray, Q. Tian, Q. Deng, D. Moseley, A-H. Chen, J. Lapano, J. S. Gardner, G. Eres, T. Z. Ward, E. Feng, H. Cao, V. Lauter, M. A. McGuire, R. Hermann, D. Parker, M-G. Han, A. Kayani, G. Rimal, L. Wu, T. R. Charlton, R. G. Moore, M. Brahlek,
Stoichiometry-Induced Ferromagnetism in Altermagnetic Candidate MnTe,
\href{https://doi.org/10.1002/adfm.202405829}
{Adv. Funct. Mater. \textbf{34}, 2405829 (2024).}

\bibitem{Hariki2024prl} A. Hariki, A. Dal Din, O. J. Amin, T. Yamaguchi, A. Badura, D. Kriegner, K. W. Edmonds, R. P. Campion, P. Wadley, D. Backes, L. S. I. Veiga, S. S. Dhesi, G. Springholz, L. \v{S}mejkal, K. V\'yborn\'y, T. Jungwirth, and J. Kune\v{s},
X-Ray Magnetic Circular Dichroism in Altermagnetic $\alpha$-MnTe,
\href{https://doi.org/10.1103/PhysRevLett.132.176701}
{Phys. Rev. Lett. \textbf{132}, 176701 (2024).}

\bibitem{Osumi2024prb} T. Osumi, S. Souma, T. Aoyama, K. Yamauchi, A. Honma, K. Nakayama, T. Takahashi, K. Ohgushi, and T. Sato,
Observation of a giant band splitting in altermagnetic MnTe,
\href{https://doi.org/10.1103/PhysRevB.109.115102}
{Phys. Rev. B \textbf{109}, 115102 (2024).}

\bibitem{Amin2024nature} O. J. Amin, A. Dal Din, E. Golias, Y. Niu, A. Zakharov, S. C. Fromage, C. J. B. Fields, S. L. Heywood, R. B. Cousins, F. Maccherozzi, J. Krempask\'y, J. H. Dil, D. Kriegner, B. Kiraly, R. P. Campion, A. W. Rushforth, K. W. Edmonds, S. S. Dhesi, L. \v{S}mejkal, T. Jungwirth, and P. Wadley,
Nanoscale imaging and control of altermagnetism in MnTe,
\href{https://doi.org/10.1038/s41586-024-08234-x}
{Nature \textbf{636}, 348–353 (2024).}

\bibitem{Smejkal2022prx} L. \u{S}mejkal, J. Sinova, and T. Jungwirth,
Emerging Research Landscape of Altermagnetism,
\href{https://doi.org/10.1103/PhysRevX.12.040501}
{Phys. Rev. X \textbf{12}, 040501 (2022).}

\bibitem{Fedchenko2024sa} O. Fedchenko, J. Min\'ar, A. Akashdeep, S. W. D'Souza, D. Vasilyev, O. Tkach, L. Odenbreit, Q. Nguyen, D. Kutnyakhov, N. Wind, L. Wenthaus, M. Scholz, K. Rossnagel, M. Hoesch, M. Aeschlimann, B. Stadtm\"uller, M. Kl\"aui, G. Sch\"onhense, T. Jungwirth, A. B. Hellenes, G. Jakob, L. \v{S}mejkal, J. Sinova, H-J. Elmers,
Observation of time-reversal symmetry breaking in the band structure of altermagnetic RuO$_2$,
\href{https://doi.org/10.1126/sciadv.adj4883}
{Sci. Adv. \textbf{10}, eadj4883 (2024).}

\bibitem{Chen2025cpl} W. Chen, L. Zeng and W. Zhu,
Helicity-Controlled Spin Hall Angle in 2D Altermagnets with Rashba Spin-Orbit Coupling,
\href{https://doi.org/10.1088/0256-307X/42/1/017201}
{Chin. Phys. Lett. \textbf{42}, 017201 (2025).}

\bibitem{Sato2024prl} T. Sato, S. Haddad, I. C. Fulga, F. F. Assaad, and J. van den Brink,
Altermagnetic Anomalous Hall Effect Emerging from Electronic Correlations,
\href{https://doi.org/10.1103/PhysRevLett.133.086503}
{Phys. Rev. Lett. \textbf{133}, 086503 (2024).}

\bibitem{Roig2025prl} M. Roig, Y. Yu, R. C. Ekman, A. Kreisel, B. M. Andersen, and D. F. Agterberg,
Quasisymmetry-Constrained Spin Ferromagnetism in Altermagnets,
\href{https://doi.org/10.1103/839n-rckn}
{Phys. Rev. Lett. \textbf{135}, 016703 (2025).}

\bibitem{Attias2024prb} L. Attias, A. Levchenko, and M. Khodas,
Intrinsic anomalous Hall effect in altermagnets,
\href{https://doi.org/10.1103/PhysRevB.110.094425}
{Phys. Rev. B \textbf{110}, 094425 (2024).}

\bibitem{Takahashi2025npj} K. Takahashi, H-F. Huang, J-X. Yu, and J. Zang,
Symmetry and minimal Hamiltonian of nonsymmorphic collinear antiferromagnet MnTe,
\href{https://doi.org/10.1038/s41535-025-00784-1}
{npj \emph{Quantum Materials} \textbf{10}, 70 (2025).}

\bibitem{Smolenski2025arxiv} S. Smolenski, N. Mao, D. Zhang, Y. Guo, A.K.M. A. Shawon, M. Xu, E. Downey, T. Musall, M. Yi, W. Xie, C. Jozwiak, A. Bostwick, N. Tamura, E. Rotenberg, L. Li, K. Sun, Y. Zhang, N. H. Jo,
Strain-tunability of the multipolar Berry curvature in altermagnet MnTe,
\href{https://doi.org/10.48550/arXiv.2509.21481}
{\emph{arXiv}:2509.21481}

\bibitem{Junior2023prb} P. E. F. Junior, K. A. de Mare, K. Zollner, K-h. Ahn, S. I. Erlingsson, M. van Schilfgaarde, and K. V\'yborn\'y,
Sensitivity of the MnTe valence band to the orientation of magnetic moments,
\href{https://doi.org/10.1103/PhysRevB.107.L100417}
{Phys. Rev. B \textbf{107}, L100417 (2023).}

\bibitem{Ren2017jmcc} Y. Ren, J. Yang, Q. Jiang, D. Zhang, Z. Zhou, X. Li, J. Xin and X. He,
Synergistic effect by Na doping and S substitution for high thermoelectric performance of p-type MnTe,
\href{https://doi.org/10.1039/C7TC01480E}
{J. Mater. Chem. C. \textbf{5}, 5076-5082 (2017).}

\bibitem{Xin2019ne} J. Xin, J. Yang, Q. Jiang, S. Li, A. Basit, H. Hu, Q. Long, S. Li, X. Li,
Reinforced bond covalency and multiscale hierarchical architecture to high performance eco-friendly MnTe-based thermoelectric materials,
\href{https://doi.org/10.1016/j.nanoen.2019.01.003}
{Nano Energy, \textbf{57}, 703–710 (2019).}

\bibitem{Devaraj2024prm} N. Devaraj, A. Bose, and A. Narayan,
Interplay of altermagnetism and pressure in hexagonal and orthorhombic MnTe,
\href{https://doi.org/10.1103/PhysRevMaterials.8.104407}
{Phys. Rev. Materials \textbf{8}, 104407 (2024).}

\bibitem{Liu2025arvix} Z. Liu, S. Xu, J. M. DeStefano, E. Rosenberg, T. Zhang, J. Li, M. B. Stone, F. Ye, R. Cong, S. Pan, C-W. Chu, L. Deng, E. Morosan, R. M. Fernandes, J-H. Chu, P. Dai,
Strain-tunable anomalous Hall effect in hexagonal MnTe,
\href{https://doi.org/10.48550/arXiv.2509.19582}
{\emph{arXiv}:2509.19582}

\bibitem{Bangar2025arxiv} H. Bangar, P. Tsipas, P. Rout, L. Pandey, A. Kalaboukhov, A. Lintzeris, A. Dimoulas, S. P. Dash,
Interplay between altermagnetic order and crystal symmetry probed using magnetotransport in epitaxial altermagnet MnTe,
\href{https://doi.org/10.48550/arXiv.2505.14589}
{\emph{arXiv}:2505.14589}

\bibitem{Belashchenko2025prl} K. D. Belashchenko,
Giant Strain-Induced Spin Splitting Effect in MnTe, a g-Wave Altermagnetic Semiconductor,
\href{https://doi.org/10.1103/PhysRevLett.134.086701}
{Phys. Rev. Lett. \textbf{134}, 086701 (2025).}

\bibitem{Zeng2006prl} C. Zeng, Y. Yao, Q. Niu, and H. H. Weitering,
Linear Magnetization Dependence of the Intrinsic Anomalous Hall Effect,
\href{https://doi.org/10.1103/PhysRevLett.96.037204}
{Phys. Rev. Lett. \textbf{96}, 037204 (2025).}

\bibitem{Takahashi2017prb} R. Takahashi and N. Nagaosa,
Berry curvature and orbital angular momentum of electrons in angle-resolved photoemission spectroscopy,
\href{https://doi.org/10.1103/PhysRevB.91.245133}
{Phys. Rev. B \textbf{91}, 245133 (2017).}

\bibitem{Nagaosa2010rmp} N. Nagaosa, J. Sinova, S. Onoda, A. H. MacDonald, and N. P. Ong,
Anomalous Hall effect,
\href{https://doi.org/10.1103/RevModPhys.82.1539}
{Rev. Mod. Phys. \textbf{82}, 1539 (2010).}

\bibitem{Yates2007prb} J. R. Yates, X. Wang, D. Vanderbilt, and I. Souza,
Spectral and Fermi surface properties from Wannier interpolation,
\href{https://doi.org/10.1103/PhysRevB.75.195121}
{Phys. Rev. B \textbf{75}, 195121 (2007).}

\bibitem{Xiao2006prl} D. Xiao, Y. Yao, Z. Fang, and Q. Niu,
Berry-Phase Effect in Anomalous Thermoelectric Transport,
\href{https://doi.org/10.1103/PhysRevLett.97.026603}
{Phys. Rev. Lett. \textbf{97}, 026603 (2006).}

\bibitem{Bergman2010prl} D. L. Bergman and V. Oganesyan,
Theory of Dissipationless Nernst Effects,
\href{https://doi.org/10.1103/PhysRevLett.104.066601}
{Phys. Rev. Lett. \textbf{104}, 066601 (2010).}



\bibitem{Behnia2016rpp} K. Behnia and H. Aubin,
Nernst effect in metals and superconductors: a review of concepts and experiments,
\href{https://doi.org/10.1088/0034-4885/79/4/046502}
{2016 Rep. Prog. Phys. \textbf{79}, 046502 (2016).}

\bibitem{Alam2023prb} M. S. Alam, A. Fakhredine, M. Ahmad, P. K. Tanwar, H-Y. Yang, F. Tafti, G. Cuono, R. Islam, B. Singh, A. Lynnyk, C. Autieri, and M. Matusiak,
Sign change of anomalous Hall effect and anomalous Nernst effect in the Weyl semimetal CeAlSi,
\href{https://doi.org/10.1103/PhysRevB.107.085102}
{Phys. Rev. B 107, 085102 (2023).}















	
\end{thebibliography}
\end{document}